\documentclass[apj]{emulateapj}
\usepackage{pstricks,amsmath,psfig}
\usepackage{apjfonts}



\newcommand{\snr}{CTA~1}
\newcommand{\psr}{RX~J0007.0+7302}
\newcommand{\asca}{{\em ASCA}}
\newcommand{\rosat}{{\em ROSAT}}

\newcommand{\xmm}{{\em XMM-Newton}}
\newcommand{\chandra}{{\em Chandra}}
\newcommand{\gam}{2EG~J0008+7307}
\newcommand{\gamm}{3EG~J0010+7309}

\begin{document}

\submitted{Accepted for Publication in ApJ}

\lefthead{COMPACT SOURCE IN CTA 1}
\righthead{SLANE ET AL.}

\title{X-ray Observations of the Compact Source in CTA~1}

\author{Patrick Slane\altaffilmark{1}, Erik R. Zimmerman\altaffilmark{2},
John P. Hughes\altaffilmark{3}, Frederick D. Seward\altaffilmark{1},
Bryan M. Gaensler\altaffilmark{1}, and Melanie J. Clarke\altaffilmark{1}}

\altaffiltext{1}{Harvard-Smithsonian Center for Astrophysics,
    60 Garden Street, Cambridge, MA 02138}
\altaffiltext{2}{Harvard College, Cambridge, MA 02138}
\altaffiltext{2}{Department of Physics and Astronomy, Rutgers, The State
University of New Jersey, Piscataway, NJ 08854-8019}

\begin{abstract}
The point source \psr, at the center of supernova remnant \snr,
was studied using the X-Ray Multi-mirror Mission (\xmm).  
The X-ray spectrum of the source is consistent with a neutron star
interpretation, and is well described by a power law
with the addition of a soft thermal component that may correspond
to emission from hot polar cap regions or to cooling emission from
a light element atmosphere over the entire star. There is evidence
of extended emission on small spatial scales which may correspond to
structure in the underlying synchrotron nebula. No pulsations are
observed.  Extrapolation of the nonthermal spectrum of \psr\ to gamma-ray
energies yields a flux consistent with that of EGRET source 
\gamm, supporting the proposition that there is a gamma-ray 
emitting pulsar at the center of \snr. Observations of the outer regions
of \snr\ with the Advanced Satellite for Cosmology and Astrophysics confirm
earlier detections of thermal emission from the remnant and show that
the synchrotron nebula extends to the outermost reaches of the SNR.

\end{abstract}


\section{Introduction}

\snr\ (G119.5+10.2) is one of the class of composite supernova remnants 
(SNRs) characterized
by the presence of pulsar wind nebulae (PWNe) at their centers.
The SNR has a large-diameter ($\sim 107$~arcmin) with low X-ray surface 
brightness (Seward, Schmidt, \& Slane 1995) and a
partial shell morphology in the radio with an apparent breakout into lower 
density material in the north (Sieber, Salter, \& Mayer 1981).
A kinematic distance of $1.4 \pm 0.3$~kpc has been derived based on the 
association of an HI shell with the SNR (Pineault et al. 1993).

\rosat\ observations of \snr\ (Seward et al. 1995) 
reveal a center-filled morphology as well
as a faint compact source \psr\ located at the peak of the central 
brightness distribution (Figure 1). \asca\ observations show 
that the diffuse central
emission is nonthermal, presumably corresponding to a wind nebula
driven by an active neutron star for which \psr\ may be the
counterpart (Slane et al. 1997 -- hereafter S97). The power law index 
of this nonthermal
emission increases with distance from the center, consistent with synchrotron
losses of particles injected from a central source, and there is also
weak evidence for a thermal component with $kT \sim 0.2$~keV,
presumably corresponding to emission from the SNR shell. 
No radio counterpart to \psr\ is 
identified in a list of compact sources by Pineault et al. (1993). 

The EGRET source \gamm\ (earlier designated \gam)
lies in the direction of \snr\
(Brazier et al. 1998), and the 95\% confidence contour for the location 
is consistent with the position of \psr. The two best-established classes
of EGRET sources are blazars and pulsars, and the lack of variability
in 10 distinct observations of \gamm\ (Brazier et al. 1998)
argues in favor of the latter interpretation (although Tompkins 1999 presents 
mild evidence for variability). 

In this paper we report on new X-ray observations of \snr\ 
from the \asca\ and \xmm\ observatories in an effort to better constrain
the nature of \psr\ as a candidate pulsar in this SNR. The observations
and data reduction are described in Section 2, and the results of the
analysis are detailed in Section 3. We conclude with a discussion of
new constraints on the nature of \psr, its relationship to \snr, and
the extension of its spectrum to $\gamma$-ray energies for comparison
with \gamm.

%%%%%%%% Figures 1 & 2 %%%%%%%%%%%%%%%%
\begin{figure*}[tb]
%\pspicture(0,10.8)(18.5,21)
\pspicture(0,10.1)(18.5,21)

\rput[tl]{0}(0.0,21.5){\epsfxsize=8.5cm
\epsffile{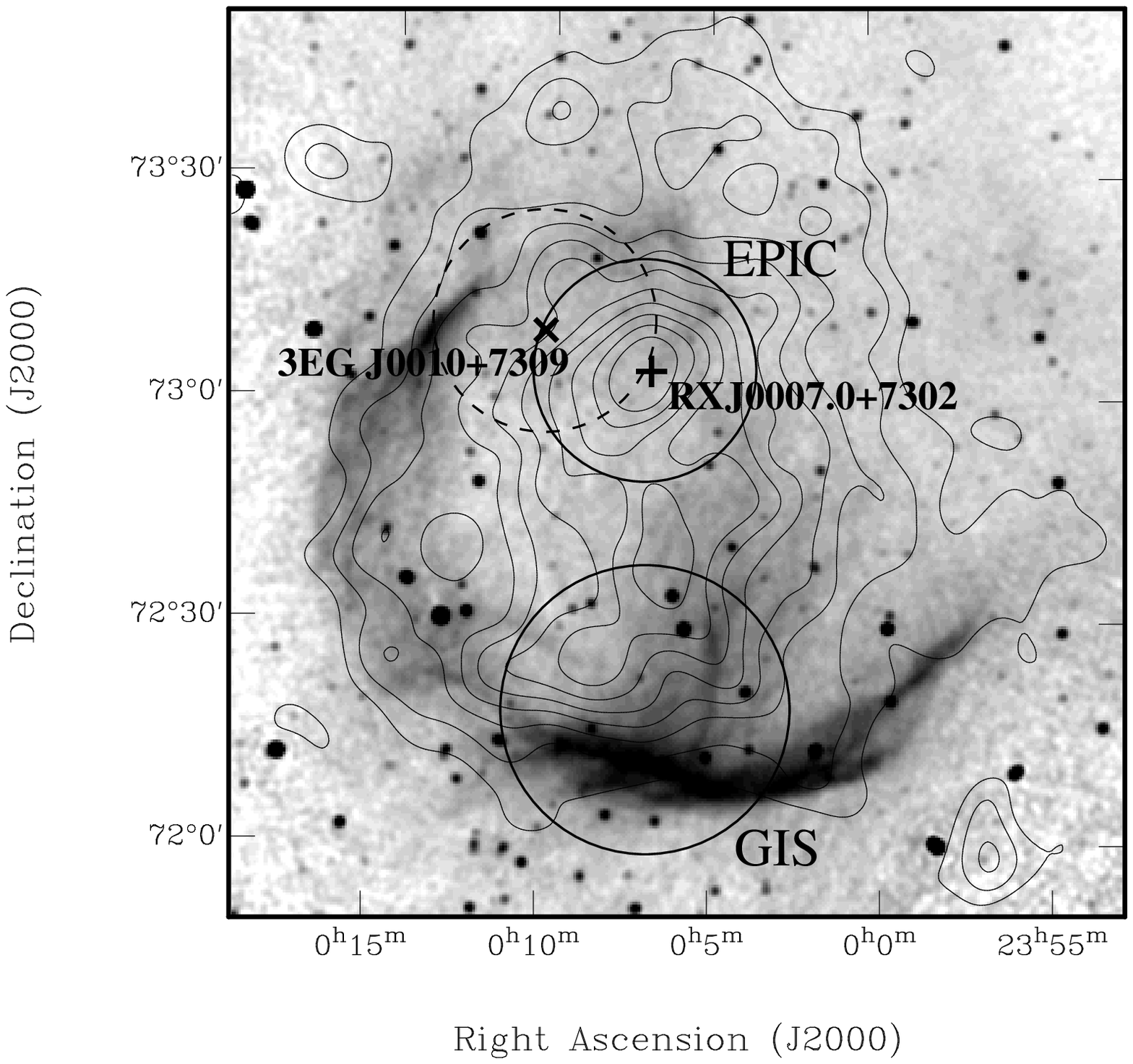}}

\rput[tl]{0}(9.75,20.7){\epsfxsize=8.0cm
\epsffile{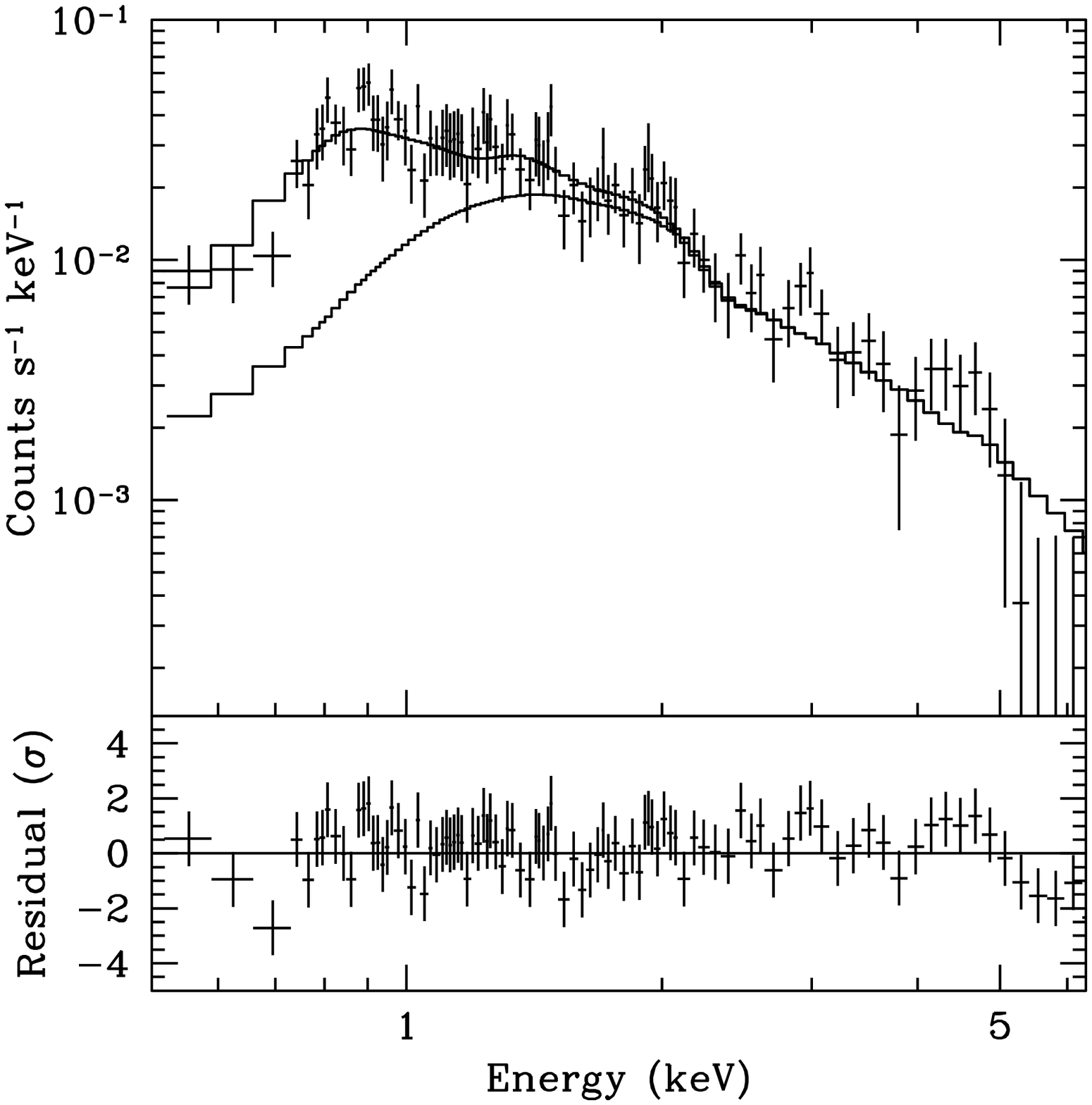}}

\rput[tl]{0}(0,12.7){
\begin{minipage}{8.75cm}
\small\parindent=3.5mm
{\sc Fig.}~1.---
Radio image of \snr\ (from Pineault et al. 1997) with PSPC contours.
The location of the \asca\ observation described here is indicated by the
GIS field of view in the southern portion of the remnant, and the \xmm\
pointing is indicated by the approximate MOS field of view in the center.
The position of \psr\ is indicated by a cross. The position of \gamm\ is
indicated by an X, with the approximate 95\% error region shown as a dashed
circle. (Note that the actual error region is slightly irregular and overlaps
with the X-ray source position.)
\end{minipage}
}

\rput[tl]{0}(9.7,12.7){
\begin{minipage}{8.75cm}
\small\parindent=3.5mm
{\sc Fig.}~2.---
\asca\ GIS2 spectrum from the southern rim of \snr. The top histogram
corresponds
to the best-fit two component model described in the text. The bottom histogram
represents only the power law component.
\end{minipage}
}

\endpspicture
\end{figure*}

\section{Observations}

\subsection{\asca\ Observations}
As an extension of our initial \asca\ studies that covered only the
central portions of \snr, we observed the southern shell of the SNR
to investigate the spectrum of the nonthermal emission far from the
remnant center, and to search for additional evidence of thermal emission
from the shell. The 60~ks observation was carried out on 10 February 1997. The 
approximate field of view of the GIS detectors is illustrated in Figure 1.
The SIS detectors were operated in 2-CCD mode, and cover a smaller portion
of the SNR with a lower effective exposure. The associated SIS spectra contain
relatively fewer counts, and we concentrate on the GIS data exclusively.

We used standard screening procedures to generate a cleaned set of events
for each detector, and extracted spectra from the region indicated
in Figure 1. Spectra were grouped to contain a minimum of 25 counts in each
spectral bin. Background spectra were extracted from the same regions of
each detector using data from the available \asca\ ``blank sky'' fields.
While these fields are at high Galactic latitude, and thus free of diffuse
emission from the Galactic Plane, they are adequate for \snr\ which lies
$\sim 10^\circ$ above the Plane.  A weighted effective area
file was generated for each extended spectral region using the 
{\tt ascaarf} task in the {\tt ftools} analysis package.
The number of counts in each
spectrum, after background subtraction, was $\sim 3000.$ The spectrum 
from GIS 2 is plotted in Figure 2 along with model fits described below.

\subsection{\xmm\ Observations}
CTA~1 was observed for 40~ks with \xmm\ on February 21, 2002 with a pointing
direction centered on \psr.  The EPIC cameras were operated in
imaging mode, with the medium and thin filters used for the two MOS detectors
and the pn detector, respectively. The small window mode with
6~ms time resolution was used for the pn detector in order to 
search for pulsed emission from the source.
Data were filtered to eliminate intervals of high background from soft
proton flares, using the Scientific Analysis Software (SAS) package, and 
events were restricted to the energy range between 100 and 15000 eV.
The resulting exposure times were $\sim 30$~ks ($\sim 26$~ks) for
the MOS (pn) detectors, and the observed count rates for \psr\ were
$1.3 \times 10^{-2} (2.9 \times 10^{-2}) {\rm\ cnt\ s}^{-1}$ in the
0.5--10~keV band.
The image of \psr\ obtained with the MOS 1 detector is shown as an inset
to Figure 4.  

Spectra of \psr\ were extracted from each EPIC detector and
regrouped to contain a minimum of 25 counts in each bin. Background spectra
were taken from regions adjacent to \psr\ in order to include similar
contributions from the diffuse emission from the synchrotron nebula.
Spectral response (RMF) and effective area (ARF) files were created for
each spectrum using the SAS tasks {\tt rmfgen} and {\tt arfgen}.
The spectrum from the MOS~1 detector is shown in Figure 3.

\section{Analysis}
\subsection{Spectral Analysis}

The \asca\ GIS spectra were modeled initially with an absorbed 
power law spectrum. The 
fit results in significant residual emission below $\sim 1$~keV (Figure 2)
which is adequately fit with an additional
component for a thermal plasma in collisional ionization equilibrium
(Raymond \& Smith 1977).
While the best-fit column density is consistent with results
from earlier \asca\ measurements (S97), 
the uncertainty is large. We thus fix this at the previously determined 
value of $N_H \sim 2.8 \times 10^{21}{\rm\ cm}^{-2}$.
The best-fit results, summarized in Table 1, yield a spectral
index of $2.3^{+0.3}_{-0.4}$ and a plasma temperature of
$kT = 0.28^{+0.3}_{-0.08}$~keV. The temperature is consistent with 
that inferred from earlier \asca\ observations. 

We first attempted to fit the EPIC spectra of \psr\ using an absorbed 
power law model. This model provides a good fit, but yields a column
density of $\sim 10^{20}{\rm\ cm}^{-2}$, which is much lower than that
for \snr. Fixing the absorption at the SNR value yields residual soft
emission which is well fit by a blackbody of $kT \sim 130$~eV with an
emitting area of radius less than 1~km. The power law index 
$\Gamma \sim 1.5$ is similar to that of the Crab Pulsar. The best-fit 
spectral parameters are listed in Table 2. The blackbody flux comprises 
roughly 20\% of the unabsorbed flux from the source. 
Replacing the blackbody component with a neutron star atmosphere model
in a $\sim 10^{12}$~G magnetic field (Pavlov et al. 1995; model kindly
provided by Slava Zavlin) reduces the temperature of the thermal component
to $T \sim 8\times 10^{5}$~K. 

%%%%%%%% Figures 3 & 4 %%%%%%%%%%%%%%%%
\begin{figure*}[t]
%\pspicture(0,10.7)(18.5,21)
\pspicture(0,10.5)(18.5,21)

\rput[tl]{0}(0.0,21.0){\epsfxsize=8.3cm
\epsffile{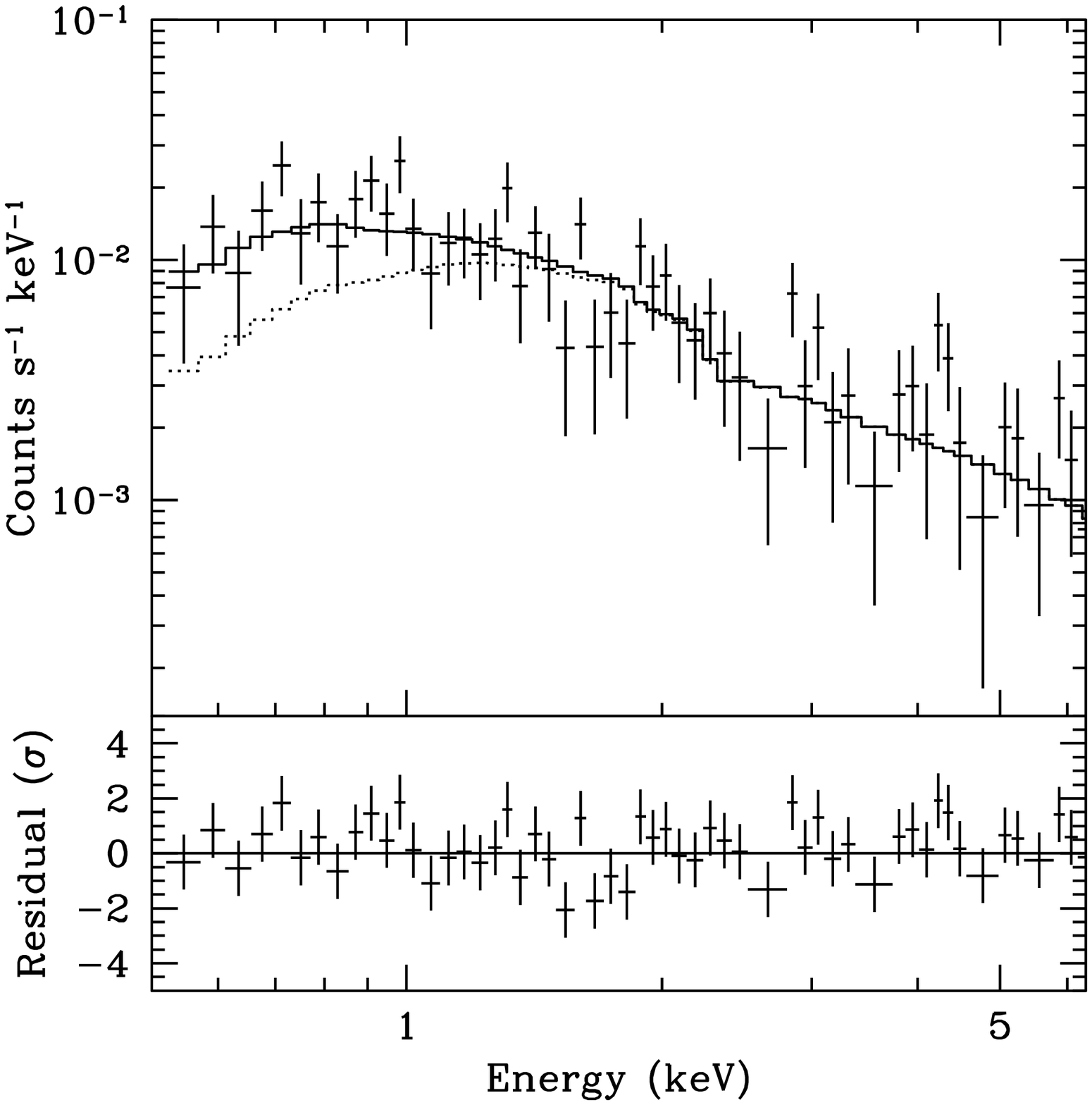}}

\rput[tl]{0}(9.75,21.0){\epsfxsize=8.0cm
\epsffile{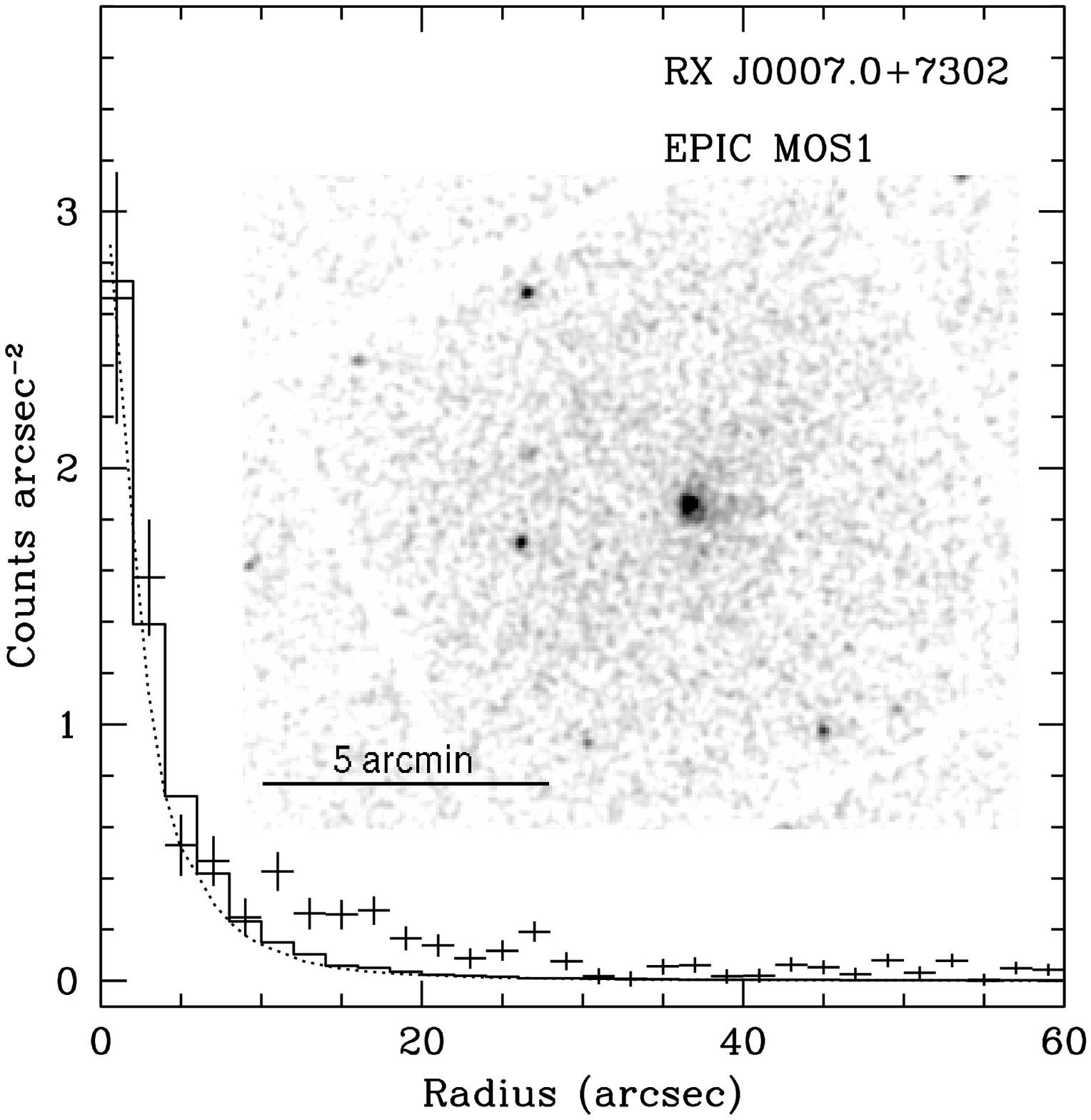}}

\rput[tl]{0}(0,12.7){
\begin{minipage}{8.75cm}
\small\parindent=3.5mm
{\sc Fig.}~3.---
\xmm\ pn spectrum of \psr\ in \snr. The upper histogram corresponds to the
power law plus neutron star atmosphere fit described in the text, while the
lower histogram represents only the power law component.
\end{minipage}
}

\rput[tl]{0}(9.7,12.7){
\begin{minipage}{8.75cm}
\small\parindent=3.5mm
{\sc Fig.}~4.---
\xmm\ MOS 1 brightness profile for \psr. Also shown is the profile for a
similar compact source in the supernova remnant G347.3--0.5 (histogram)
as well as the predicted profile for a point source (dashed curve).
There appears to be evidence for extended emission from \psr. The
MOS~1 + MOS~2 image is shown as an inset, with a scale bar at the lower
left. \psr\ is the brightest source, near the center of the image.
\end{minipage}
}

\endpspicture
\end{figure*}
%%%%%%%%%%%%%%%%%%%%%%%%%%%%%%%%%%%%%%%

\subsection{Spatial Analysis}

The nominal position of \psr\ derived from the EPIC observations is 
${\rm RA}_{2000} = 00^{\rm h}07^{\rm m}02.2^{\rm s}$, 
${\rm Dec}_{2000} = +73^\circ 03^\prime 07^{\prime \prime}$, with an
uncertainty of $\sim 4^{\prime \prime}$. 
We find evidence that the source may be slightly extended. In Figure 4 
we plot the radial brightness profile extracted from the MOS~1 data, and
compare it to that for the point source in G347.3--0.5, obtained from 
archival \xmm\ data. The point sources were located
at the center of their respective fields, and the two profiles have 
been normalized to yield the same number of counts within a 15~arcsec 
radius. Also plotted is the radial count distribution for the on-axis 
point spread function (PSF) of the MOS1 telescope in the 0.75-2.25~keV 
energy range profile, from the \xmm\ Users' Handbook (Issue 2.1), normalized 
in the same manner.
The profile for \psr\ appears to exceed that from the comparison profiles
beyond a radius of $\sim 10$~arcsec. This may
be an indication of substructure in the inner portions of the extended
nebula. Such structure is now known to be commonplace in PWNe, as dramatically
illustrated by high resolution X-ray images of the Crab and Vela pulsars,
for example (see Slane 2003). 
We note that the spectrum for \psr\ is harder than that for
the point source in G347.3--0.5 (Lazendic et al. 2003), while the 
energy-dependent point spread function (PSF) of the \xmm\ mirrors is smaller
for higher energies. The apparent extent is thus not an artifact of the
telescope PSF.

To further investigate this issue, we extracted spectra from smaller regions
surrounding \psr\ and repeated the spectral analysis. We find that the ratio
of the blackbody flux to the powerlaw flux increases for the smaller regions,
as would be expected if the powerlaw component is extended while the blackbody
component is pointlike. This result is not highly significant, however, and
observations with higher spatial resolution will be required to confirm such 
structure in \snr. Such observations have recently been carried out with
\chandra\ (PI J. Halpern).

\subsection{Timing Analysis}

We used the EPIC pn data to search for pulsed emission from \psr. The 
SAS task {\tt epchain} was used to reprocess events to correct a timing
error inherent in some \xmm\ data, and events were then corrected 
to the solar system barycenter with the task {\tt barycen}, using the 
coordinates derived above. A total of 1055 counts in the energy band
0.3-10~keV were extracted from within a radius of 9.6~arcsec, centered on 
\psr, with an estimated $483 \pm 34$ originating from the source 
(the rest being background events). A Fast-Fourier Transform was applied to the 
binned light curve using $\sim 8.4 \times 10^6$ bins spanning frequencies 
up to 83.3~Hz. No significant evidence of pulsations is observed in the 
resulting power spectrum. The maximum power in the FFT (aside from values 
corresponding to the $\sim 40$~ks duration of the observation) was 28.4, from 
which we derive an upper limit of 61\% for the pulsed fraction of a sinusoidal 
profile in the quoted frequency range (Vaughan et al. 1994).

\section{Discussion}

Under the Sedov solution (Sedov 1959), 
assuming a shell-like morphology similar to the radio emission,
with a nominal radius of 40~arcmin, and 
extrapolating our spectral results to the entire SNR, we infer an age of
$1.3 \times 10^4 D_{1.4}$~yr (where $D_{1.4}$ is the distance in units of
1.4~kpc), and an ambient density of $1.7 \times 10^{-2}
D_{1.4}^{-1/2}{\rm\ cm}^{-3}$ assuming that the GIS field covers roughly
15\% of the emission from the shell. These values, while rather uncertain
given the small region observed, are in good agreement with estimates based 
on earlier \asca\ observations (S97) and indicate a typical SNR that has 
evolved in a somewhat low density environment. We note that the density
value corresponds to the southern region covered by the ASCA observation;
the northern region, which extends beyond the nominal radius used here,
presumably corresponds to a lower density.

The new \asca\ observations show that the diffuse synchrotron emission in
\snr\ extends to the outer boundaries of the SNR, at least in the
southern shell region. Although the telescope
resolution does not allow us to identify the outer extent with precision,
it is clear that the radius of this component is larger than the $\sim 18$
arcmin ($7.2 D_{1.4}$ pc) extent estimated in S97. 
The PWN is quite large relative to the size of the surrounding shell,
although larger PWNe are known (Gaensler, Dickel, \& Green 2000).
The spectral (photon) index 
in the outer regions of the nebula may be slightly steeper than the central 
value of $2.1 \pm 0.1$, but not significantly so. This suggests little 
synchrotron
aging of the X-ray emitting electrons, which is indicative of a low
magnetic field. This is similar to what is observed for the PWN powered
by PSR~B1509--58 (Gaensler et al. 2002).

The spectral results derived for \psr\ are consistent with the interpretation
that this source is the pulsar that powers the surrounding synchrotron
nebula. The luminosity of the power law component from the compact
source is $L_x = 4.7
\times 10^{31} d_{1.4}^2 {\rm\ erg\ s}^{-1}$. This is low for a pulsar
young enough to have an X-ray luminous SNR, but not exceedingly so: 
PSR~B1853+01 is only a factor of two brighter, and is also associated
with an observed SNR (W44). The luminosity of the PWN in \snr\ 
is $5.6 \times 10^{33} d_{1.4}^2 {\rm\ erg\ s}^{-1}$ (S97)
which means the point source comprises roughly 1\% of the nonthermal
emission, a value similar to that for pulsars in 3C~58, G54.1+0.3,
and G292.0+1.8. Assuming a spin-down power $\dot E \sim 10^3 L_x
\approx 5 \times 10^{34} d_{1.4}^2 {\rm\ erg\ s}^{-1}$ from 
the crude correlation derived from X-ray studies of other pulsars 
(Seward \& Wang 1988, Becker \& Tr\"umper 1999), \psr\ apparently has 
sufficient energy to power the nebula.
The derived spectral index for \psr\ is similar to that of the Crab Pulsar.
If the source is indeed a pulsar with the inferred $\dot E$, this would
differ considerably from the empirical relationship between spectral 
index and spin-down power suggested by Gotthelf (2003).

The position of \psr\ is offset from the geometrical center of the SNR (defined
here by the circular portion of the radio shell, ignoring the blowout region in
the north). Using the age above, this offset corresponds to a transverse
velocity of $\sim 450 {\rm\ km\ s}^{-1}$ which is reasonable for a pulsar.
We note, however, that the apparent nonuniform density of the surrounding ISM
introduces considerable uncertainty in estimating the center of the SNR.

%%%%%%%%%%%%%%%%%% TABLE 1 %%%%%%%%%%%%%%%%%%%%%%%%%%%%%
\begin{center}
Table 1: Spectral Parameters
\scriptsize
\begin{tabular}{cccc}\\ \hline
& \snr & \multicolumn{2}{c}{\psr} \\
& \asca\ GIS &
\multicolumn{2}{c}{\xmm\ EPIC} \\

& PL + Thermal & PL + BB & PL + NSA \\ \hline

$N_H ({\rm\ cm}^{-2}$) &&
$2.8 \times 10^{21}$ (fixed) \\ \hline

$\Gamma$ (photon) & $2.3^{+0.2}_{-0.4}$ &
$1.5 \pm 0.2$ & $1.6 \pm 0.2$\\

$F_x ({\rm\ erg\ cm}^{-2}{\rm\ s}^{-1})^a$ & $4.8 \times 10^{-12}$ &
$2.0 \times 10^{-13}$ & $2.0 \times 10^{-13}$ \\ \hline

$kT$ (keV) & $0.28^{+0.3}_{-0.08}$ &
$0.136 \pm 0.012$ & $0.053 \pm 0.004$ \\

$F_x ({\rm\ erg\ cm}^{-2}{\rm\ s}^{-1})^a$ & $1.1 \times 10^{-11}$ &
$3.3 \times 10^{-14}$ & $3.5 \times 10^{-14}$ \\

$R$(km) & & $0.63 D_{1.4}$ & 10 (fixed) \\ \hline

$\chi^2/dof$ & 437.2/474 & 117.5/116 & 112.2/116 \\ \hline
\\
\multicolumn{3}{l}{a) Unabsorbed flux (0.5-10 keV)} \\
\end{tabular}
\end{center}
\normalsize
%%%%%%%%%%%%%%%%%%%%%%%%%%%%%%%%%%%%%%%%%%%%%%%%%

The X-ray spectral index of $\sim 1.5$ is similar to that for
\gamm\ in $\gamma$-rays ($1.6 \pm 0.2$). We find that extrapolation
of the spectrum for \psr\ using $\Gamma = 1.4$ predicts a flux of
$4 \times 10^{-7}{\rm\ photons\ cm}^{-2}{\rm\ s}^{-1}$ for $E > 100$~MeV, in
good agreement with the value measured by EGRET
($4.6 \pm 0.6 \times 10^{-7}{\rm\ photons\ cm}^{-2}{\rm\ s}^{-1}$). Thus,
the spectrum of \psr\ is capable of powering the EGRET flux of \gamm\
with no spectral break between the two energy bands. This, combined with the
fact that there is no brighter X-ray source in the EGRET error box, and that
both \psr\ and \gamm\ possess properties indicative of a pulsar,
strongly suggests that these sources are one and the same.

The temperature of the blackbody component in the spectrum of \psr\ is too
high, and the emitting area is too low, for this to be interpreted as
cooling from the surface of a young neutron star
(e.g. Kaminker et al. 2002). Rather, this component
could be associated with the heating of polar cap regions of
the neutron star by instreaming particles accelerated in the magnetosphere
(e.g. Pavlov et al. 2000, Zavlin et al. 2002).
The lack of observed pulsations from such an emission structure is not
worrisome at the upper limit levels derived here. Strong
gravitational bending of the emission from near the surface of the NS
can easily reduce the pulsed fraction to such levels (e.g. Psaltis, 
\"Ozel, \& DeDeo 2000).
More sensitive timing measurements are obviously of considerable
interest in an effort to detect pulsations that could potentially secure
the identification between \psr\ and \gamm, as well as provide
the spindown properties of the pulsar. Deep radio observations have yet
to reveal pulsations, with a flux density upper limit of 0.8~mJy at
606~MHz (Lorimer et al. 1998), although recent observations have uncovered
much fainter pulsars in the young SNRs G292.0+1.8 (Camilo et al. 2002a)
and 3C~58 (Camilo et al. 2002b), both of which are also faint pulsed
X-ray sources (Hughes et al. 2003, Murray et al. 2002).
Deep X-ray timing observations of \psr\ are currently planned with the 
\chandra\ HRC (PI S. Murray).

Alternatively, the thermal emission may be the result of cooling from the
entire NS surface, viewed through a light element atmosphere. The inferred
temperature for such a model, described above, is below that expected for
a NS whose age is 10-20~kyr (as estimated for the SNR) assuming cooling
by the modified Urca process, but consistent with models in which the
direct Urca process becomes active in sufficiently massive stars
(Kaminker et al. 2002, Yakovlev et al. 2002). Such models would predict
a mass of $\sim 1.44 M_\odot$ for \psr.

\section{Conclusions}
We have used \asca\ and \xmm\ data to study \snr\ and the apparently associated
compact source at its center, \psr. The diffuse nonthermal emission extending
from the central regions of the SNR is consistent with previous 
interpretations of \snr\ as a composite SNR containing a large PWN.
The faint thermal emission detected along the southern shell 
is consistent with expectations for a moderate age SNR expanding into
a low density environment, as might be encountered at the high Galactic
latitude of \snr.

The nonthermal nature of the spectrum from \psr\ strongly suggests
that this source is a NS with an actively emitting magnetosphere,
and the existence of a blackbody-like component in the spectrum indicates
the presence of either hot polar caps or cooling of the neutron star surface
through an atmosphere.
The X-ray luminosity of the source is reasonable for an active pulsar,
though rather low for one of such a young age. We do not detect pulsations
from the source, and place an upper limit of 61\% on the pulsed fraction,
which is lower than that for some known X-ray emitting pulsars, but
is not overly restrictive on the interpretation of this as the
pulsar powering the synchrotron nebula in \snr.

We find evidence that the power law component from \psr\ may be 
extended, suggesting that there could be inner structure to the 
known large PWN, and indicating that the pulsar is still supplying energy
to the larger nebula.  This inner structure could be similar to 
the jets or toroidal structures that have been recently identified for several 
pulsars using high-resolution \chandra\ observations.  

Finally, we find additional evidence for a connection between \psr\ and the 
EGRET source \gamm\ due to the similarities in the spectra of the two 
sources. This strongly, though not conclusively, suggests that
\psr\ is a pulsar that is emitting both X-rays and $\gamma$-rays.
More sensitive searches for pulsed emission from the source, through deeper
X-ray observations as well as radio observations, are of considerable
interest in an effort to confirm this picture.

\acknowledgements{We thank Jan Vrtilek and Dima Yakovlev for helpful 
discussions related to this study, Jasmina Lazendic for help with processing
the \xmm\ data, and Slava Zavlin and George Pavlov for providing the NS
atmosphere model.  POS acknowledges support from NASA
contract NAS8-39073 and grants NAG5-4803 and NAG5-10017.}

\end{document}